\documentclass[aps,prb,superscriptaddress,twocolumn,floatfix]{revtex4}
\usepackage{graphicx}
\usepackage{epstopdf}
\usepackage[]{amsmath}
\usepackage{gensymb}

\newcommand{\YTO}{Yb$_2$Ti$_2$O$_7$ }
\newcommand{\YTOns}{Yb$_2$Ti$_2$O$_7$}
\newcommand{\YTOd}{Yb$_{2+x}$Ti$_{2-x}$O$_{7-\delta}$ }
\newcommand{\YTOdns}{Yb$_{2+x}$Ti$_{2-x}$O$_{7-\delta}$}
\newcommand{\YTOs}{Yb$_{2+x}$Ti$_{2-x}$O$_{7-x/2}$ }

\newcommand{\YTOdef}{Yb$_2$TiO$_5$ }

\begin{document}

\title{Impact of Stoichiometry of \YTO on its Physical Properties}

\author{K. E. Arpino} 
\affiliation{Institute of Quantum Matter, The Johns Hopkins University, Baltimore MD 21218 USA}
\affiliation{Department of Chemistry, The Johns Hopkins University, Baltimore MD 21218 USA}

\author{B. A. Trump} 
\affiliation{Institute of Quantum Matter, The Johns Hopkins University, Baltimore MD 21218 USA}
\affiliation{Department of Chemistry, The Johns Hopkins University, Baltimore MD 21218 USA}

\author{A. O. Scheie} 
\affiliation{Institute of Quantum Matter, The Johns Hopkins University, Baltimore MD 21218 USA}
\affiliation{Department of Physics and Astronomy, The Johns Hopkins University, Baltimore MD 21218 USA}

\author{T. M. McQueen} 
\affiliation{Institute of Quantum Matter, The Johns Hopkins University, Baltimore MD 21218 USA}
\affiliation{Department of Chemistry, The Johns Hopkins University, Baltimore MD 21218 USA}
\affiliation{Department of Physics and Astronomy, The Johns Hopkins University, Baltimore MD 21218 USA}
\affiliation{Department of Materials Science \& Engineering, The Johns Hopkins University, Baltimore MD 21218 USA}

\author{S. M. Koohpayeh} 
\altaffiliation{Corresponding author: koohpayeh@jhu.edu}
\affiliation{Institute of Quantum Matter, The Johns Hopkins University, Baltimore MD 21218 USA}
\affiliation{Department of Physics and Astronomy, The Johns Hopkins University, Baltimore MD 21218 USA}
\email{koohpayeh@jhu.edu}

\date{\today}

\newcommand{\chem}[1]{\ensuremath{\mathrm{#1}}}

\begin{abstract}
A series of \YTOd doped samples demonstrates the effects of off-stoichiometry on \YTOns's structure, properties, and magnetic ground state via x-ray diffraction, specific heat, and magnetization measurements. A stoichiometric single crystal of \YTO grown by the traveling solvent floating zone technique (solvent~=~30~wt\%~rutile TiO$_2$ and 70~wt\%~\YTOns) is characterized and evaluated in light of this series. Our data shows that upon positive $x$ doping, the cubic lattice parameter $a$ increases and the Curie-Weiss temperature $\theta_{CW}$ decreases. Heat capacity measurements of stoichiometric \YTO samples exhibit a sharp, first-order peak at $T$ = 268(4)~mK that is suppressed in magnitude and temperature in samples doped off ideal stoichiometry. The full entropy recovered per Yb ion is 5.7 ~J~K$^{-1}~\approx~Rln2$. Our work establishes the effects of doping on \YTOns's physical properties, which provides further evidence indicating that previous crystals grown by the traditional floating zone method are doped off ideal stoichiometry. Additionally, we present how to grow high-quality colorless single crystals of \YTO by the traveling solvent floating zone growth method.
\end{abstract}


\maketitle

\section{Introduction}
Materials with the pyrochlore structure ($A_2B_2$O$_7$) are a topic of extensive study in the field of magnetism as ideal hosts for prototypical geometric frustration including both classical and quantum spin-ice and spin-liquid behavior.\cite{GeoFrust,MagPyrochlores,SpinIceReview,SpinLiquid,SpinLiquidRev} In this structure type, the $A$ and $B$ metal ions each form a sublattice of a corner-sharing tetrahedra; the two sublattices are inter-penetrating. Ising-like (uniaxial) spin interactions on either of these geometrically frustrated tetrahedra sublattices can give rise to spin-ice behavior.\cite{GeoFrust,SpinIceReview} In the case of \YTOns, the magnetic behavior originates in the Yb\textsuperscript{3+} (4$f$\textsuperscript{13}) ions which make up the $A$ sublattice. In an ideally stoichiometric sample of \YTOns, the magnetic behavior of this Yb\textsuperscript{3+} sublattice is isolated from any interfering magnetic interactions originating in the interpenetrating $B$ sublattice because 3$d$\textsuperscript{0} Ti\textsuperscript{4+} has no valence electrons. 

\YTOns's magnetic interactions may be described by an anisotropic exchange Hamiltonian whose exchange parameters have been experimentally ascertained by multiple groups.\cite{Ross_2011_PRX,Applegate_2012,Hayre_2013,Robert_2015,Coldea} Based on the determined exchange parameters, the ground state of \YTO has been predicted to lie near phase boundaries between ferromagnetic and antiferromagnetic states in theorized phase diagrams,\cite{Ross_2011_PRX,Chang_2012,Wong_2013,Yan_2013,Coldea,Jaubert_2015,Robert_2015,Chen_2016} and it has been suggested that the quantum fluctuations resulting from proximity to these phase boundaries could potentially make \YTO a quantum spin liquid candidate.\cite{Ross_2011_PRX} \YTO has been considered a quantum spin ice candidate due to a finding that the predominant spin exchange is ferromagnetic along the local axis of the tetrahedra\cite{Ross_2011_PRX,Applegate_2012,Hayre_2013}; however, recent reports indicate the ground state is not that of a quantum spin ice.\cite{Jaubert_2015,Robert_2015} A number of experimental and theoretical investigations into the true nature of \YTOns's ground state described it in widely varying and conflicting fashions: either as lacking\cite{Hodges_2002,Yaouanc_2003,Gardner_2004,Bonville_2004,Ross_2009,Hayre_2013,DOrtenzio_2013,Bhattacharjee_2016}
 or having\cite{Yasui_2003,Chang_2012,Applegate_2012,Chang_2014,Lhotel_2014,Gaudet_2016}
  long-range magnetic order, with static\cite{Chang_2014, Gaudet_2016}
 to slowly fluctuating dynamic\cite{Hodges_2002,Yaouanc_2003,Gardner_2004,Bonville_2004,Cao_2009_JPCM,Ross_2011_PRB,DOrtenzio_2013,Lhotel_2014} spins.

Experimental inconsistencies in both the \YTO samples themselves and the measured low-temperature transition to the ground state hinder identification of the true ground state. While powder samples are generally white, the color of crystals varies from deep red to translucent yellow-gray.\cite{FZGrowth_2013,RossThesis} \YTO has a transition to the presumed ground state at $T~\sim$~250 mK which appears to vary considerably between samples. The specific heat signatures of powders generally have a sharp, intense peak around $T$~=~260~mK,\cite{Physica_1969,Dalmas_2006,Ross_2011_PRB,DOrtenzio_2013,Chang_2014} while those of single crystals grown by the traditional floating-zone method have peaks which occur at lower temperatures ($T \sim$~150 to 200~mK) and are reduced in height by an order of magnitude.\cite{Ross_2011_PRB,Yaouanc_2011,Chang_2012,DOrtenzio_2013,Chang_2014} Similarly, low-temperature magnetization measurements exhibit transitions at a higher temperature in powder samples than in single-crystal samples ($T =$~245~mK compared to  $T =$~150~mK, respectively).\cite{Lhotel_2014} These differences suggest a systematic material discrepancy between single crystal and powder samples, and indicate a better and more consistent method of crystal growth is needed.

Off-stoichiometry likely plays an important role in understanding these differences: in similar pyrochlores, physical properties have been shown to be highly dependent on sample stoichiometry.\cite{PZO, HTO,TTO, TTO_2} In \YTOns, one crystal has been found to be Yb-doped, or ``stuffed'' (having Yb  on the Ti-site), despite having been made from purely stoichiometric powder.\cite{Ross_2012,Gaudet_2015} Additionally, Chang et al. observed diminished intensity of EXAFS data for those crystals with lower heat capacity peaks in a study of three crystals and concluded that Yb deficiency is responsible for broadening the heat capacity peak.\cite{Chang_2012} On this basis, it has been generally assumed that all \YTO single crystal are non-stoichiometric, and this off-stoichiometry explains the discrepancy that has been observed between crystal and powder data. However, to our knowledge, the correlation between the stoichiometry and physical properties of \YTO which underlies this assumption has not be systematically investigated.

In this paper, we report the structural and physical properties characterization of  a polycrystalline \YTOd doped series to elucidate the effects of off-stoichiometry on structure, heat capacity and entropy, and magnetic susceptibility. Additionally, a colorless single crystal of apparently stoichiometric \YTO grown by the traveling solvent floating zone (TSFZ) technique is characterized alongside the doped series. The  low-temperature specific heat of the series displays a dramatic change in breadth, height, and temperature of the transition upon doping, while the single crystal has a single, notably sharp peak at $T$~=~268(4)~mK.

\section{Experimental procedures}
\label{section:EP}
\subsection{Synthesis}
Synthesized powders having target stoichiometries \YTOs ($x$~= 0.08, 0.02, 0.01, 0.00, -0.01, and -0.02) were prepared from precursors Yb$_2$O$_3$ (99.99\% Alfa Aesar) and rutile TiO$_2$ (99.99\% Alfa Aesar) in large ($\sim$20~g) batches to minimize mass error. Prior to use, precursors were dried at 1200~\degree C overnight. Amounts of precursors used were calculated on the basis of metal ion stoichiometries under the assumption that oxygen concentration, either excess or deficient, would be corrected by heating in ambient atmosphere (\YTOdns). Precursors were combined and intimately ground in a porcelain mortar and pestle, then loaded into an alumina crucible and heated under ambient atmosphere to 1200~\degree C and held at temperature for 12~h. The material was removed, intimately ground, then pressed into a pellet and reheated in an alumina crucible under ambient atmosphere to 1350~\degree C and held at temperature for 10~h. This step was repeated several times until powder x-ray diffraction (XRD) patterns showed the precursors were fully reacted. The obtained powders were white.

Sintered rods (rigid, polycrystalline samples) were prepared by compacting and pressing synthesized powders into the form of cylindrical rods (of approximately 6~mm in diameter and 70-80~mm in length), and then sintering at higher temperatures in a four-mirror optical floating zone furnace (Crystal System Inc. FZ-T-4000-H-VII-VPO-PC equipped with four 1-kW halogen lamps) under 2~atm O$_2$. These polycrystalline rods were sintered by a single pass zone heating below the melting point:  a power level of 62\% was used for all but the $x~=~0$ sample, which was heated at 68\% lamp power.

 A pure, stoichiometric single crystal approximately 5 mm in diameter and 40 mm  in length was obtained via the TSFZ technique \cite{SK_2008,SK_2016} using a four-mirror optical floating zone furnace (Crystal System Inc. FZ-T-4000-H-VII-VPO-PC) equipped with four 1-kW halogen lamps as the heating source. The feed and seed rods, attached to the upper and lower shafts respectively, were sintered rods of stoichiometric \YTO powder (above). The seed rod had been used in a previous \YTO growth; any residual crystalline material at its top (the crystal growth base) was sanded flat. The $\sim$0.2 g solvent pellet used was composed of 30\% rutile TiO$_2$ and  70\% stoichiometric \YTO by mass. The solvent pellet was first melted and joined to the feed and seed rod in the optical furnace before beginning the growth. During the growth, the molten zone was passed upwards at a rate of 0.5 mm/h. Rotation rates of 3 and 6 rpm were employed in opposite directions for the feed rod (upper shaft) and the growing crystal (lower shaft), respectively. Crystal growth was carried out at a power level of 64.2\%, which remained fixed throughout the growth, under a dynamic oxygen atmosphere with a pressure of 1 atm and a flow rate of 10 mL/min. Only one zone pass was performed in a crystal growth. Additional clear \YTO single crystals of similar size were grown using the same parameters, indicating reproducibility.

\subsection{Characterization}
 Powder X-ray diffraction (XRD) patterns were taken using a Bruker D8 Focus X-ray diffractometer operating with Cu K$\alpha$ radiation and a LynxEye detector. Diffraction data was analyzed using the Bruker TOPAS software (Bruker AXS). For consistency, all the refinements reported in this paper were done on scans ranging from $2\theta$ = 5 to 120~\degree{} with silicon added to the sample as an internal standard. The use of a silicon standard is especially important in cubic materials such as this one, as cubic materials have only one internal lattice parameter which can convolute with the measurement parameters. Uncertainties and error bars reported for lattice parameters reflect the statistical error unless otherwise noted. Site-mixing and doping were individually tested for by holding all other parameters constant from a sample's best fit and intentionally varying the Yb and Ti occupancies to plot the Rwp as a function of site-mixing or doping; errors were estimated from the resulting plot via Hamilton R-ratio tests and visual inspection of the fits.

Physical property characterization was performed using a Quantum Design Physical Properties Measurement System (PPMS). Magnetization data were collected using the ACMS option at $T$ = 2-300~K under $\mu_0H$ = 0.2~T and converted to magnetic susceptibility using the approximation $\chi = M/H$. Curie-Weiss analyses were performed by adjusting $\chi_0$ to achieve the most linear $(\chi-\chi_0)^{-1}$ in for the temperature range $T$ = 2-30~K. Heat capacity data were collected at constant pressure in two ways: using the semiadiabatic pulse technique (2\% heat rise) for $T$ = 0.1-2~K and using a large heat pulse (100-200\% heat rise from the base temperature, or a 100-200~mK pulse) in the $T$ = 0.1-0.4~K  range of the peak. In the semiadiabatic pulse method, $C_p$ is assumed to be constant over a single short measurement and is extracted by fitting a $2\tau$ heat flow model to the temperature-heat curve. In the large heat pulse method, $C_p$ is not assumed to be constant over a longer measurement, but have distinct values in a series of $\delta{}T$ bins. Removing the constant-$C_p$ assumption is known to be better able to capture heat capacity accurately in a first-order transition. Differences in the data between the two methods result from applying heat pulses of different sizes over a transition with considerable latent heat and hysteresis. For consistency, temperature and heat capacity numbers were taken from heating traces for the large heat pulse data, which showed some hysteresis between heating and cooling traces (the peak in the cooling curve appeared approximately 5~mK lower than in the heating curve).

The entropy was calculated by integrating heat capacity divided by temperature of the sintered series up to $T$ = 2~K and single crystal up to $T$ = 300~K. For all samples, the heat capacity was assumed to be zero at $T$ = 0~K in order to place the entropy on an absolute scale, though only entropy for measured temperatures is plotted in Figure~\ref{fig:Entropy}. This assumption does not affect $\Delta{}S$. In the range of $T$~=~5~to~30~K, the magnetic entropy was isolated by subtracting the heat capacity of isostructural, non-magnetic Lu$_2$Ti$_2$O$_7$ to remove the lattice contribution.\cite{FZGrowth_2013} Below $T$ = 5~K, no lattice contribution was subtracted because the heat capacity of Lu$_2$Ti$_2$O$_7$ becomes negligibly small (e.g., C $<$ 0.01~J~K$^{-1}$ (mol magnetic ion)$^{-1}$ up to $T$ = 2~K, which is less than our \YTO measurement error). As the Debye temperatures of the two pyrochlores should be more than 99.5\% similar due to atomic masses and stoichiometry\cite{ATari}, the heat capacity of Lu$_2$Ti$_2$O$_7$ was not scaled before subtracting it from that of \YTOns. The entropy curves are only negligibly different for data taken by different heat-pulse techniques.

\section{Results and discussion}

\subsection{Structure}
Laboratory x-ray diffraction (XRD) patterns of the \YTOd synthesized powders and sintered rods samples with target stoichiometries \ $x$ = 0.08, 0.02, 0.01, 0.00, -0.01, and -0.02 were analyzed via  Rietveld refinements using a cubic pyrochlore model to extract lattice parameter, metal-ion substitution, and site-mixing. Doping was presumed to occur as Yb\textsuperscript{3+} and Ti\textsuperscript{4+} ions substituting on the other's Wyckoff positions, with the overall charge discrepancy accommodated by oxygen vacancies or interstitials. Due to oxygen's relatively low x-ray scattering factor in this compound, only the occupancy of the metal ions could be reliably modeled. The XRD patterns show a pyrochlore structure: the presence of peaks such as the (111) in XRD patterns rule out a disordered fluorite structure which is possible for some $A_2B_2$O$_7$ materials.\cite{Cava_2006}

\begin{figure}
\centering
\includegraphics[width=8.5cm]{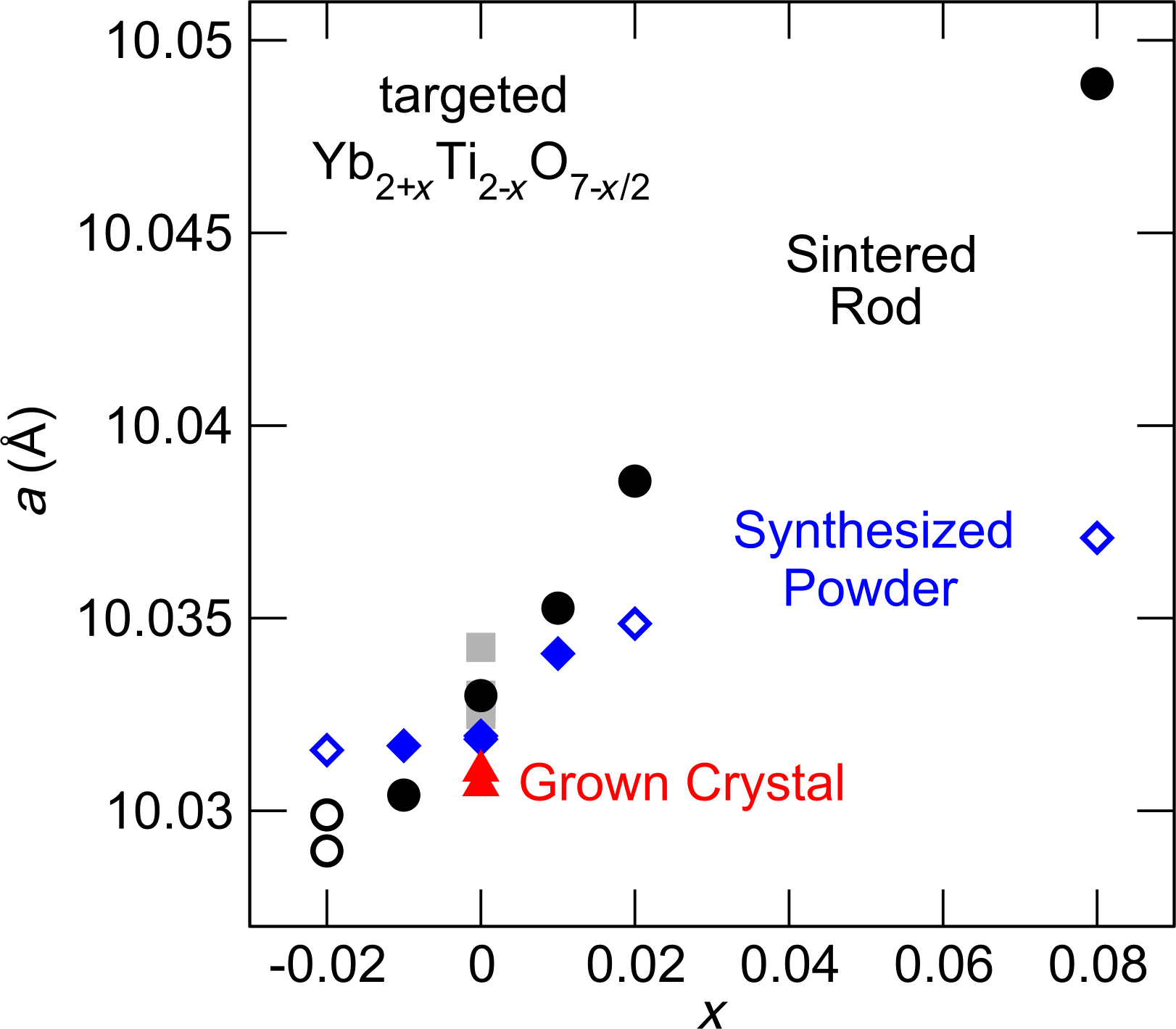}
\caption{The cubic lattice parameter $a$ of the synthesized powder (blue diamonds) and sintered rods (black circles) of the \YTOd series is shown, along with that of the grown single crystal (red triangles, $x$ estimated by Rietveld refinement) and literature values (gray squares).\cite{Lian_2003,Cava_2006,FZGrowth_2013,Cava_2015} Samples with impurity phases are depicted as hollow symbols: the  $x = 0.02$ and $x = 0.08$ synthesized powder samples have a secondary \YTOdef phase, and all $x = -0.02$ samples have a secondary phase of TiO$_2$. Error bars are contained within the symbols.}
\label{fig:LP}
\end{figure}

The cubic lattice parameters $a$ of synthesized powders and sintered rods \YTOd samples are plotted in Figure ~\ref{fig:LP}, along with those of single crystal samples. Stoichiometric ($x$ = 0) samples agree with most literature values\cite{Lian_2003,Cava_2006,FZGrowth_2013,Cava_2015}, although some studies found the lattice parameter at ambient temperatures to be notably lower, around 10.025 \AA{}.\cite{Ross_2012,Bramwell_2000} As noted in \ref{section:EP}, cubic lattice parameters are especially sensitive to experimental parameters.

XRD patterns show secondary phases in the synthesized powder samples at higher targeted levels of doping. In the $x$ = 0.02 and 0.08 synthesized powder samples, a Ti-deficient \YTOdef secondary phase\cite{Cava_2006} was observed, indicating the samples have not achieved their targeted stoichiometries, likely due to narrow \YTO phasewidth at lower temperatures as predicted in the composition-temperature phase diagram.\cite{PD} Within the limits of our x-ray diffractometer, this impurity phase was not seen in the corresponding doped sintered rod samples, which were processed at higher temperature than synthesized powders. This likely indicates a  wider phasewidth towards positive $x$ doping at higher temperatures (above 1350~\degree{}~C), which facilitated achievement of the targeted stoichiometries. In the $x$ = -0.02  samples, however, both the synthesized powder and the sintered rod samples show rutile TiO$_2$ ($\sim$1-3\% TiO$_2$ by mass) as a secondary phase, possibly indicating a limited phasewidth towards negative $x$ doping, consistent with the phase diagram.\cite{PD}

For both the synthesized powder and sintered rod \YTOd series, lattice parameter $a$ increases with $x$, as expected because Yb\textsuperscript{3+} is larger than Ti\textsuperscript{4+}. The slope of this trend is lesser for the synthesized powder samples, which could result from the samples having a smaller magnitude of doping than intended (due to incomplete reaction and/or side products) or from site-mixing of the cation sites, which essentially averages the ion size. We observe both these effects in Rietveld analysis of the synthesized powders: we detail above the presence of additional phases in higher-doped samples of the synthesized powder series (TiO$_2$ for Ti-rich $x < 0$ and \YTOdef for Yb-rich $x >0$), and testing for site-mixing in Rietveld refinement reveals approximately 1\% site mixing in synthesized powder samples (site-mixing for these six samples refines in the range of 0.5 to 1.2\% site mixing with error averaging to 0.35\%). In contrast, the sintered rod samples show no evidence of site mixing in Rietveld analysis, refining to zero site mixing as the clear minimum in the goodness of fit. Note that due to the atomic numbers of Yb and Ti (70 and 22, respectively), it is impossible to robustly distinguish whether the $x >$ 0 samples have site-mixing or Yb vacancies based on laboratory X-ray diffraction data.

\subsection{Single crystal}
\label{subs:sxtal}
Initial attempts to grow a single crystal of \YTO at its melting point via the previously reported floating zone method\cite{Yasui_2003,Ross_2009,Yaouanc_2011,Chang_2012,FZGrowth_2013} from pressed rods of polycrystalline, stoichiometric \YTO (no flux/solvent) yielded a single crystal with a number of questionable features. Firstly, the crystal was dark red in color, which is unexpected for Yb\textsuperscript{3+} and Ti\textsuperscript{4+} ions: Yb\textsuperscript{3+} absorbs in the UV, while Ti\textsuperscript{4+} has no valence electrons to absorb visible light. Density functional theory (DFT) calculations of the bandstructure support this analysis by predicting an ultraviolet 3.34 eV bandgap.\cite{Gap} Secondly, the cubic lattice parameter $a$ revealed a significant gradient along the grown crystal which varied from $a$~=~10.05147(5) \AA{} at the start of the grown crystal, consistent with extremely Yb-stuffed, to $a$~=~10.03201(7) \AA{} towards the end, consistent with nearly stoichiometric. Finally, XRD analysis of the frozen remnants of the molten zone showed evidence of titanium oxides as well as \YTO with a smaller lattice parameter ($a$~=~10.0280(1) \AA, consistent with $x < 0$ doping). These features suggest that \YTO melts incongruently, in which case the TSFZ method offers an avenue for crystal growth.\cite{SK_2008,SK_2016} Due to the existence of a TiO$_2$--\YTO eutectic at $T~=$~1620~\degree~C, it is possible based on the phase diagram \cite{PD} to use a TiO$_2$--\YTO flux to lower the temperature of the molten zone and precipitate out solid \YTOns. The specifics of our \YTO TSFZ crystal growth are given in Section~\ref{section:EP}.

\begin{figure}
\centering
\includegraphics[width=8.5cm]{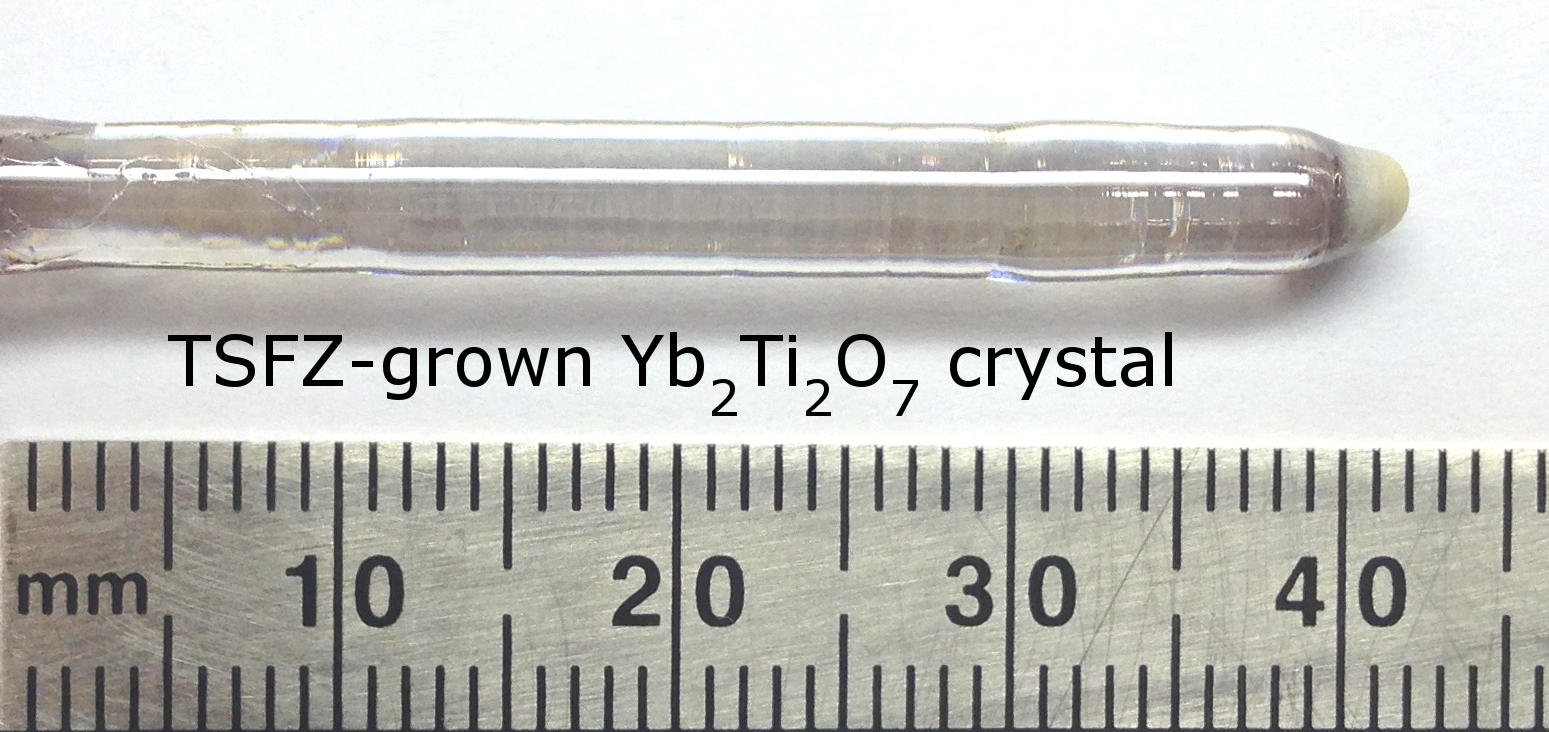}
\caption{The TSFZ technique (solvent = 30 wt\% TiO$_2$ and 70 wt\% \YTOns) produces a large single crystal of \YTO that is clear and colorless.}
\label{fig:xtal}
\end{figure}

Using the TSFZ method, the large (40~mm in length, 4~mm in diameter) clear single crystal pictured in Figure~\ref{fig:xtal} was obtained. The quality and purity of this crystal is supported by its color, steady growth temperature, and Rietveld analysis. As stated above, a stoichiometric crystal of \YTO ought to be colorless due to the ions and the predicted bandgap.\cite{Gap} It was not necessary to adjust the lamp power level during the growth: it was kept precisely at 64.2\% for the whole crystal growth. The power level of an optical furnace correlates with the growth temperature, which is highly sensitive to any slight changes in the composition of the molten zone. The steady lamp power during the growth therefore indicates an unchanging stoichiometry of the molten zone: the material leaving the molten zone (the grown crystal) is identical in composition to the material entering the molten zone (the stoichiometric feed rod).\cite{SK_2008,SK_2016} The quality and purity of this crystal were analyzed by Rietveld refinement to the XRD data. The lattice parameter does not change appreciably over the 40 mm length of the grown crystal: $a$ =10.03066(6)~\AA\ at the start and 10.03104(2) \AA\ at the end. Refinements specifically to check for \YTOs doping and site-mixing indicate the crystal may be possibly Ti-rich, refining at $x=-0.005(9)$ with no (0.0(4)\%) site mixing indicated. The possibility of small amount of excess Ti is consistent with the crystal growth method and the slightly lowered lattice parameter; however, the metal-ion ratio is within error of stoichiometric. The crystal therefore appears to be of high quality and approximately stoichiometric based on structural analysis; analysis of the physical properties (discussed below) agrees with this conclusion.

\subsection{Heat capacity and entropy}

\begin{figure}
\centering
\includegraphics[width=8.5cm]{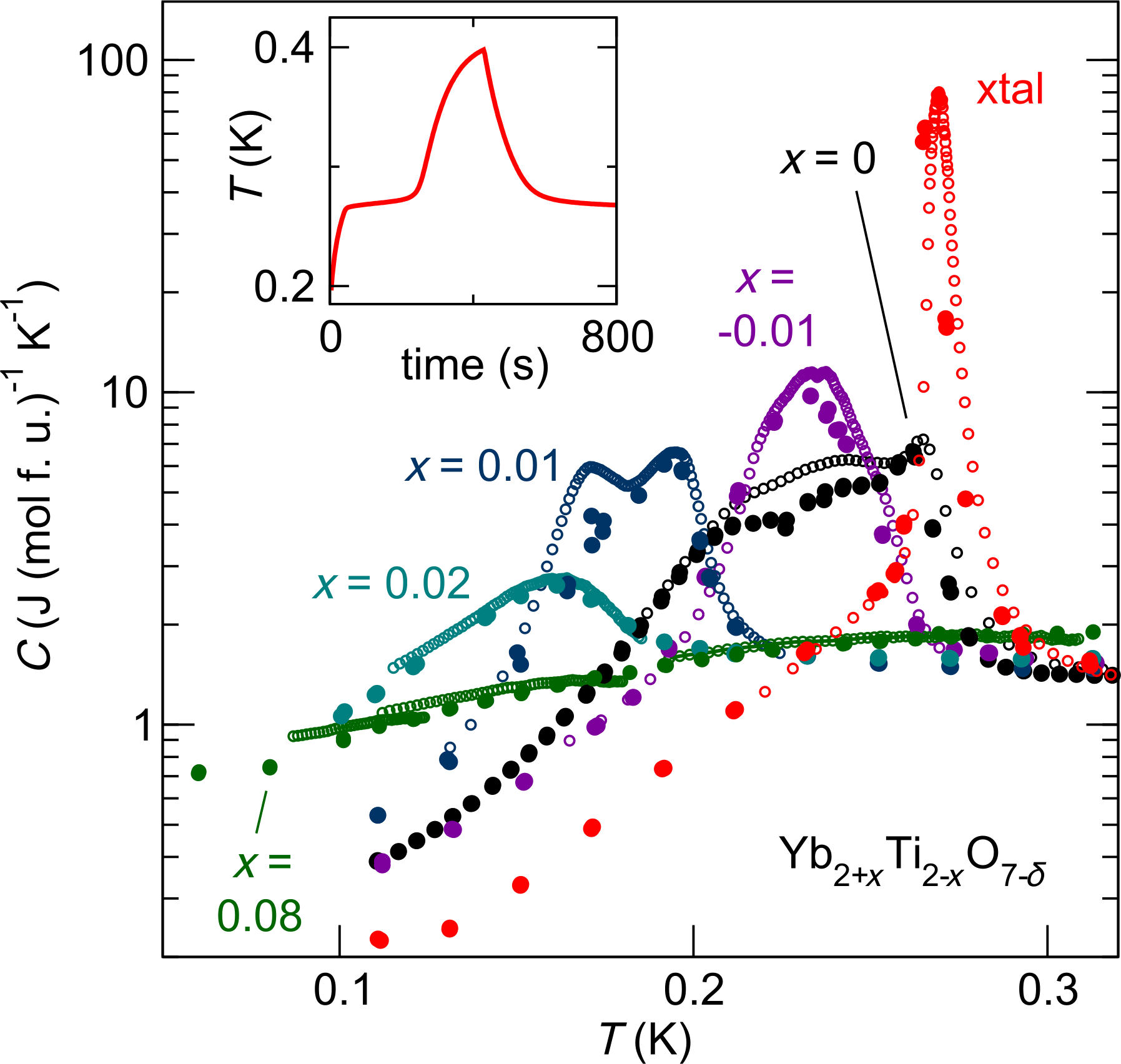}
\caption{The specific heat capacity of the single crystal (red) and sintered rod \YTOd series (darker colors, labeled) exhibit a peak at $T < 300$ mK which broadens and decreases in temperature as \YTOd is doped away from $x = 0$. Data taken by the semi-adiabatic method are plotted as filled circles, and data taken by the large heat pulse method are plotted as smaller empty circles. Inset: heating trace of the single crystal taken by the large heat pulse method shows a first-order transition, as evidenced by a plateau in the temperature at around $T \sim$~270~mK.}
\label{fig:HC}
\end{figure}

Low-temperature ($T$ = $\sim$~0.1 to 2 K) heat capacity measurements taken on the single crystal and on the sintered rod series are shown in Fig. \ref{fig:HC}. The specific heat of the single crystal exhibits a single, large (87(9) J (mol f. u.$)^{-1}$ K $^{-1}$), sharp peak at $T$ = 268(4)~mK with no other notable features. This is the only crystal in the known literature whose heat capacity shares these attributes with powder data\cite{Dalmas_2006,Ross_2011_PRB, DOrtenzio_2013,Chang_2014}; a table of heat capacity signatures from the literature is given in the Appendix for comparison. Given that there is evidence that at least one literature single crystal is off-stoichiometric\cite{Ross_2012} and a general belief that the off-stoichiometry of single crystals is the cause of the different heat capacity signature, this may suggest our single crystal is not doped off-stoichiometry like literature crystals are presumed to be. A large amount of latent heat is visible in the heat trace of the long pulse data (Fig. \ref{fig:HC} inset). The latent heat and the sharpness of the heat capacity peak suggest the transition is first order; additionally, there appears to be some hysteresis in the precise temperature at which the peak occurs depending on whether the data is taken upon heating or cooling, with the temperature of the peak in the cooling curve being about 5~mK lower.

Our sintered rod \YTOd series displays a trend of the heat capacity peak broadening and decreasing in $T$ and $C$ upon doping. Even slight ($x = \pm0.01$) doping greatly changes the temperature at which the peak is centered, from $T \sim268$~mK to $T \sim235$ and $\sim194$~mK. The peak appears to become increasingly broader with doping. For the $x = 0.08$ sample, any feature which may be present is so broad we do not observe it as a recognizable peak, but only in the increased specific heat at the lowest temperatures compared to the other samples.

Our stoichiometric sintered rod sample exhibits a narrow initial peak at $T$ = 265~mK along with at least one broad feature in the $200<T<250~mK$ range. Though somewhat obscured by its placement on the logarithmic scale, this initial feature spikes over 1 J (mol f. u.$)^{-1}$ K $^{-1}$)  higher than the surrounding heat capacity and  is less than 10 mK wide. It is close in temperature to the heat capacity peak of our single crystal and is similar in temperature, breadth, and height to the sharp initial feature of several crystals by Ross et al. which also have an initial sharp feature and broad humps at lower temperature.\cite{Ross_2011_PRB,Ross_2012} We hypothesize that the features below the sharp initial peak in our sample are due to included off-stoichiometry phases. Note a higher sintering temperature was used for this sample (68\% lamp power vs. 62\% used for the other rods); any partial melting of the sample during the sintering process may have formed non-stoichiometric phases if \YTO is an incongruent melter as is suggested in Section~\ref{subs:sxtal}.

 The low, broad peak in heat capacity data of our intermediately doped samples, e.g., the $x = 0.02$, appears most similar to the features of single crystals in the literature: although there is some variety in the specifics of  peak shape for the various single crystal samples in the literature (namely the ``sharpness'' of the peak), all share a decreased and broadened peak centered at lower $T$ than stoichiometric powder data. Single crystals in the literature have a broad peak of C $\leq$ 5~J (mol Yb)$^{-1}$ K $^{-1}$ in the $T$ = 150-200 mK temperature range, \cite{Yaouanc_2011,Ross_2011_PRB,Chang_2012,DOrtenzio_2013,Chang_2014} while powders have much larger peaks at higher temperature ($T \sim260$~mK)\cite{Physica_1969,Dalmas_2006,Ross_2011_PRB,DOrtenzio_2013,Chang_2014}; see the Appendix for specifics. Considering the series of doped samples presented in this work, we posit that the crystals in the literature are not stoichiometric, but instead are doped off-stoichiometry systematically due to the crystal growth method employed. Ross et al. have already found one of these single crystals to be doped at a level of $x = 0.046(4)$.\cite{Ross_2012} Our series of doping levels illustrates explicitly that the heat capacity peak decreases in both height and temperature in doped \YTOns. 

\begin{figure}
\centering
\includegraphics[width=8.5cm]{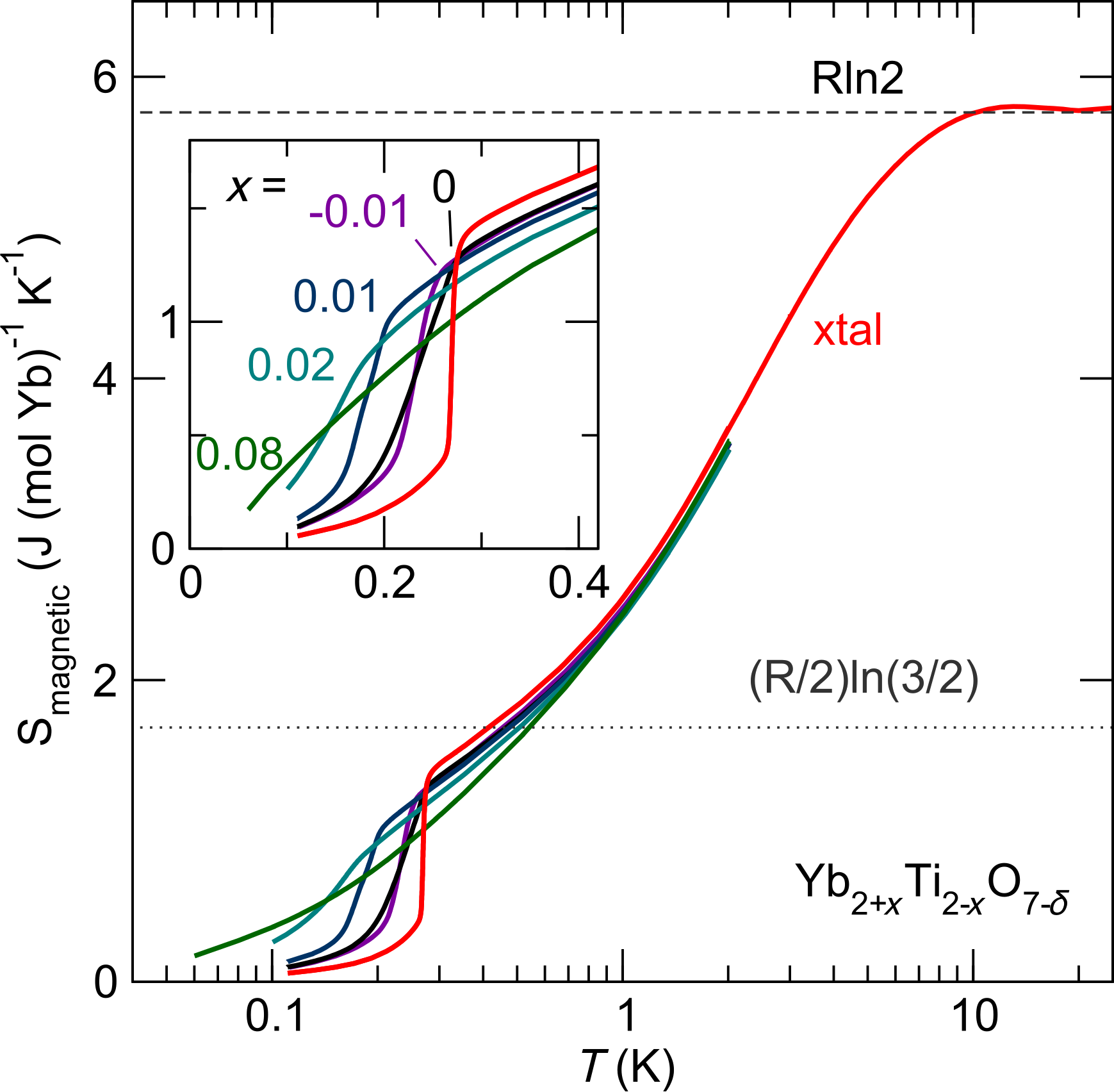}
\caption{Magnetic entropy per mole Yb of the single crystal (red) integrated over $T$~=~$\sim$~0.1 to 20~K levels off at the two-state entropy $R$ln2 rather than the spin ice limit $Rln2-(R/2)ln(3/2)$. Magnetic entropy of the sintered rod \YTOd series (darker colors, labeled in insert) integrated over $T$~=~$\sim$~0.1 to 2~K show that the sharp decrease in entropy at  the $T \sim270$~mK transition broadens significantly as \YTOd is doped away from $x = 0$ (detailed view in inset). Entropy associated with this broadened transition persists in the doped samples to lower temperatures.}
\label{fig:Entropy}
\end{figure}

Integrating the heat capacity over temperature with the lattice contribution removed as described in Section \ref{section:EP} allows evaluation of the magnetic entropy, shown in Fig. \ref{fig:Entropy}. The magnetic entropy of the stoichiometric single crystal shows a sharp increase at $T \sim270$~mK corresponding to the transition as well as a significant increase in the $T$ = 2 to 10 K range corresponding to a hump in the heat capacity, which indicates significant spin correlations above the magnetic transition. The total magnetic entropy per mole Yb appears to approach the two-state spin entropy $Rln2$. The entropy recovered clearly exceeds the spin-ice entropy of $Rln2-(R/2)ln(3/2)$, confirming that the magnetic ground state of \YTO cannot be that of a spin ice.\cite{Jaubert_2015,Robert_2015} The amount of entropy recovered in the transition is similar to the residual entropy of spin-ice; however, the lack of plateau observed indicated the existence of a stable spin-ice-like state above the transition is unlikely. Our results are consistent with \YTO powder and crystalline results in the literature,\cite{Chang_2014,Applegate_2012,Physica_1969} while there is some conflicting reports on the total entropy recovered for similar spin-ice pyrochlores.\cite{DTO,HTO,Ramirez_1999} Recovering the full $Rln2$ magnetic entropy in total is consistent with the ground state of \YTO being a  ferromagnetic splayed ice state\cite{Gaudet_2016,Yaouanc_2016} or collinear ferromagnetic state,\cite{Yasui_2003,Chang_2012} either of which would not have macroscopic residual entropy. Note that as $\Delta{}S = 5.72$ ~J~K$^{-1}$ (mol Yb)$^{-1}$ for 0.11~K$\leq T \leq $20~K, the assumption that C($T$~=~0~K)~$=0$ does not qualitatively affect the result that the entropy recovered is $Rln2$ per mole Yb ion.

Our doped \YTOd samples all appear to recover similar total entropy as the single crystal over the transition. While these samples were only measured up to $T$ = 2~K, the magnetic entropy at $T$ = 2~K of doped \YTOd is comparative to that of the stoichiometric crystal (C~$\sim$~3.5~J~K$^{-1}$ (mol Yb)$^{-1}$), and it is reasonable to assume the doped samples approach the same $Rln2$ per mole Yb limit as the crystal. The recovery of entropy is broadened out over a range of temperatures, especially for the $x = 0.08$ sample. This broad recovery of entropy, in conjunction with a lower transition temperature, indicates spin rearrangement associated with the transition extends to very low temperature in doped samples (e.g., $T$ = 60~mK in $x = 0.08$ sample). This could indicate that measurements of dynamic spins at low temperatures\cite{Hodges_2002,Yaouanc_2003,Gardner_2004,Bonville_2004,Cao_2009_JPCM,Ross_2011_PRB,DOrtenzio_2013,Lhotel_2014} may not be accurate measurements of the true ground state if taken on an inadvertently doped sample, as spin entropy associated with the transition remains to lower temperatures for doped samples such as the traditionally grown single crystals.

\subsection{Magnetic susceptibility}
\begin{figure}
\includegraphics[width=8.5cm]{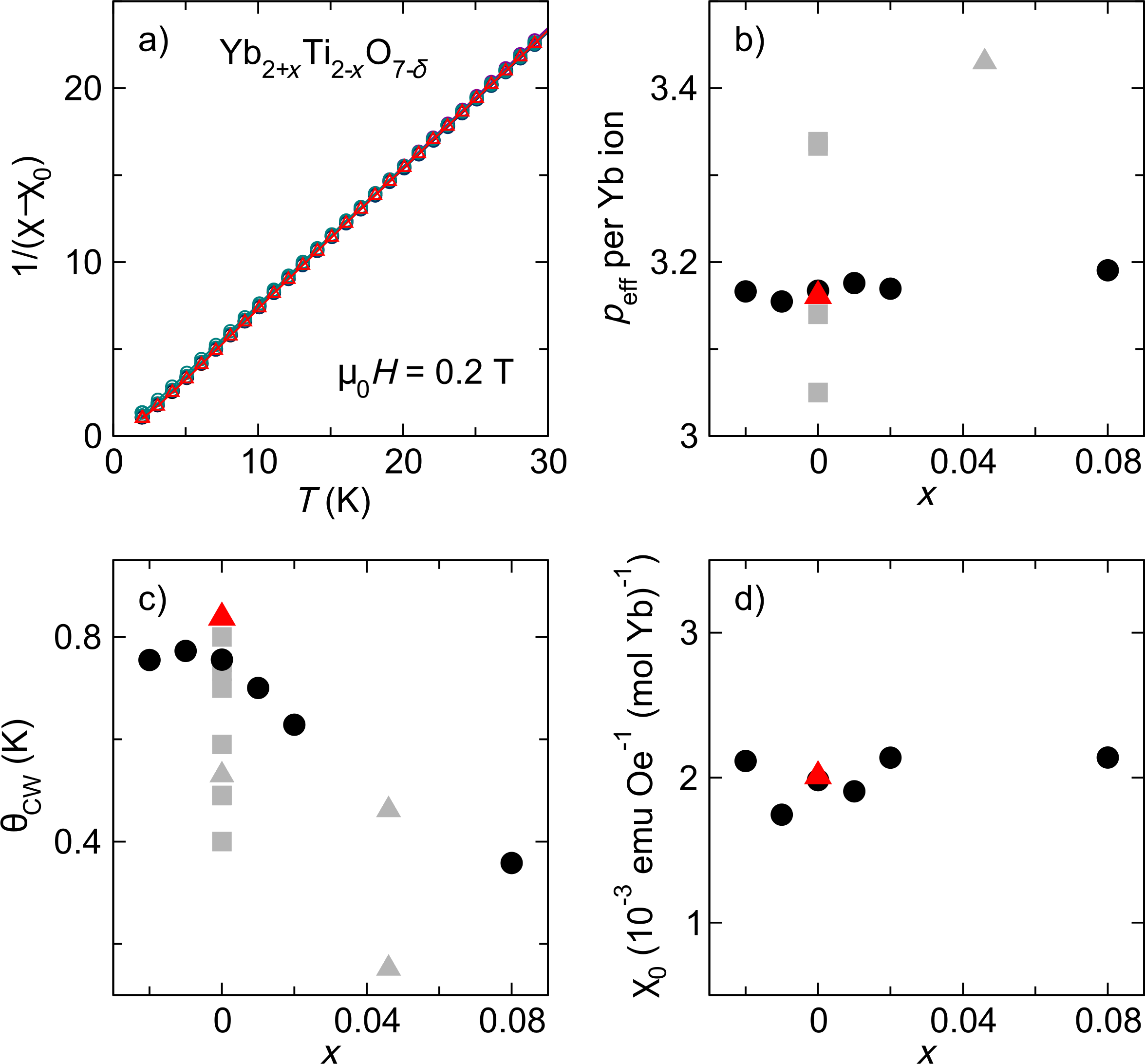}
\caption{Curie-Weiss analysis of the magnetic susceptibility of the single crystal (red triangles in b-d) and the \YTOd sintered rod series (black circles in b-d), with literature values (gray squares and triangles in b-c denote powder\cite{Physica_1969,Bramwell_2000,Hodges_2001,Cao_2009_JPCM,Ross_2012} and crystal samples\cite{Yasui_2003,Ross_2012}, respectively) shown for comparison. Statistical error bars from the fit are contained within the symbols. (a) Curie-Weiss fits in the $T$ = 2-30~K range appear linear. The fit to the crystal data is in red and fits to the sintered rod series are in darker colors used in Figures \ref{fig:HC} and \ref{fig:Entropy}; the fits are sufficiently similar that they overlap. (b) The effective magnetic moment $p_{\text{eff}}$ per Yb\textsuperscript{3+} ion are randomly distributed around 3.171(8), within the range of literature values. (c) The Curie-Weiss temperature decreases upon $x > 0$ doping (stuffing). (d) Temperature-independent term $\chi_0$ appears randomly distributed.}
\label{fig:Mag}
\end{figure}

Magnetization measurements for temperatures ranging from $T$ = 2 to 300~K at $\mu_0H$~=~0.2~T were performed on the \YTOd sintered rod series and the single crystal. The inverse susceptibility is non-linear over the full range even when including a $\chi_0$ term to account for small, temperature-independent contributions (e.g., weak diamagnetism from core electrons). Such non-linearity could result from occupation of excited crystal field levels\cite{1968,Bramwell_2000,Cao_2009_JPCM,Ross_2012}; however, when applied over a small range at low temperatures to avoid excited crystal field levels, the Curie-Weiss analysis can approximate a line reasonably well. Such an analysis was performed on a  per mole Yb ion basis from $T$ = 2 to 30~K to focus on the paramagnetic behavior and to best compare to literature analysis (Figure ~\ref{fig:Mag}). Analysis on a per mole Yb ion basis is  warranted by our initial assumption that substitutional doping does not change the oxidation state of the metal elements in this compound.

The effective magnetic moments $p_{\text{eff}}$ per mole Yb for the series and the crystal (Figure ~\ref{fig:Mag}b) appear to be roughly constant around 3.17, which corresponds to a Curie constant of 1.26 emu K Oe\textsuperscript{-1} per mole Yb ion. These values are similar to the literature values ($p_{\text{eff}}\approx~$3-3.5).\cite{Bramwell_2000,Hodges_2001,Ross_2012}
This result lends support for our initial assumption that Ti substitutes for Yb\textsuperscript{3+} as Ti\textsuperscript{4+} (Ti$^{\bullet}_{\text Yb}$)  rather than Ti\textsuperscript{3+} (Ti$^{\text x}_{\text Yb}$), because magnetically active 3$d^1$ Ti\textsuperscript{3+} would increase the magnetic response of the $x < 0$ samples, which is not observed. 

The Curie-Weiss temperature ($\theta_{CW}$, Figure ~\ref{fig:Mag}c)  for the whole series is small and positive ($\theta_{CW}$ = 0.7557(5) K for the stoichiometric), suggesting predominantly ferromagnetic interactions. Previous analyses find similar small, positive values for the Curie-Weiss temperature.\cite{Physica_1969,Bramwell_2000,Hodges_2001,Yasui_2003,Ross_2012,Cao_2009_JPCM,Chang_2012} There is a sharp, notable decrease in the Curie-Weiss temperature with increasing $x > 0$ (corresponding with more Yb\textsuperscript{3+} ions) down to 0.358 K. Presumably, this effect arises from $x$ Yb\textsuperscript{3+} ions substituting on the $B$ sites influencing the interactions of the corner-sharing network of Yb\textsuperscript{3+} on $A$ sites. With the caveat that quantitative comparison of parameters extracted depends considerably on the analysis parameters (including linear regression and temperature range used), we observe that the Curie-Weiss temperature of crystals in the literature appears to generally be lower than that of powders in the literature (e.g., 0.462~K for the crystal vs. 0.733~K for the powder when similarly analyzed \cite{Ross_2012}). This observation lends additional support to the hypothesis that crystals grown in the literature are off-stoichometric.\cite{Ross_2012}

The distribution of $\chi_0$  (Figure ~\ref{fig:Mag}d)  appears to be random in the range 0.0017 - 0.0022 emu Oe\textsuperscript{-1} per mole Yb ion for the series and single crystal, and likely reflects the error in the measurement technique rather than any sample dependence. In comparison, we estimate $\chi_0$ to be approximately 0.005 emu Oe\textsuperscript{-1} per mole Yb from a simulation of  $\chi^{-1}$ as a function of temperature calculated using the crystal field levels given by Gaudet et al.\cite{Gaudet_2015}

\section{Conclusions}
Here we demonstrate the effect of doping \YTO on its properties by synthesizing and analyzing a \YTOd series covering a range of stoichiometries. Notably, we show that lattice parameter the $a$ increases with doping $x$, that the Curie-Weiss temperature decreases upon stuffing ($x > 0$), that the peak in the heat capacity shifts to lower temperatures and becomes weaker and broader with doping, and that doping does not affect the total magnetic entropy recovered but does result in dynamic spins persisting to lower temperatures. The properties of crystals in the literature grown via the traditional floating zone method, specifically lower transition temperatures, broader peaks in the heat capacity, and lower Curie-Weiss temperatures, are consistent with our results for off-stoichiometric \YTOns.  It has been generally assumed that off-stoichiometry explains the discrepancy that has been observed between crystal and powder data based on limited data from a few crystals\cite{Ross_2012,Chang_2012}; our study provides the systematic experimental evidence linking stoichiometry to its effects on physical properties.
 
By nature of their structure, pyrochlores have number of nearly degenerate ground states; it is possible in such a material for even slight disorder to have significant implications. Such extreme sensitivity to deviations from perfect crystalline order is expected and observed in related pyrochlores such as Ho$_2$Ti$_2$O$_7$, Tb$_2$Ti$_2$O$_7$, and Pr$_2$Zr$_2$O$_7$ where the single ion ground state is not protected by Kramer's degeneracy.\cite{HTO,TTO, TTO_2,PZO} It is interesting that we observe a similar sensitivity in \YTO despite Yb$^{3+}$ having a single ion ground state doublet protected by Kramer's degeneracy; the force which causes such sensitivity is unknown. It is possible that the sensitivity of \YTO to disorder is due to proximity of the ground state to different phases, as predicted by several phase diagrams.\cite{Ross_2011_PRX,Chang_2012,Wong_2013,Yan_2013,Coldea,Jaubert_2015,Robert_2015,Chen_2016}

In-depth analysis of structural disorder is necessary to shed further light on the mechanisms by which doping affects the physical properties in \YTOns. Pyrochlores are known to be susceptible to a variety of structural disorders, ranging from point defects such as cationic site mixing, stuffing, and oxygen vacancies, to extended defects such as antiphase domains (sections of $A_2B_2$O$_7$ occuring as $B_2A_2$O$_7$)\cite{APDpyro}, nanoscale phase separation of doping levels\cite{PZO}, and structural instabilities\cite{YNbO, Cava_2008}. High-resolution synchrotron and neutron diffraction on \YTOd \cite{Cava_2006,Cava_2008,Ross_2012,Cava_2015,Gaudet_2015,Trump} can determine the disorder as a function of doping  and synthesis temperature, from which the mechanisms by which minimal doping has such a large effect can be posited. For example, Ross et al.\cite{Ross_2012} observe stuffed Yb ions, which could broaden the transition by producing a local strain field\cite{Gaudet_2015} that locally modulates magnetic interactions, giving rise to micro-regions with varied transition temperatures. Moreover, the ground state phase diagram could be mapped as a function of disorder, and intentional doping could tune the ground state.

In addition to characterizing the effects of doping \YTO on its physical properties, we present a new method of growing stoichiometric \YTO single crystals using the TSFZ method with a solvent mix of  \YTO and rutile  TiO$_2$. This method yields a large, pure, colorless single crystal of stoichiometric \YTO with no observable structural gradient. Diffraction, magnetic susceptibility, and heat capacity all support the quality of this crystal; specifically, the heat capacity shows a single, sharp, peak at $T$ = 268(4)~mK with a latent heat, indicative of a first order phase transition. This development is especially important due to the suspected off-stoichiometry of floating-zone grown crystals in the literature. High-quality, undoped stoichiometric crystals are necessary to determine the magnetic interactions of \YTO and its response to applied fields because doped samples may have excess interfering magnetic moments, as indicated broadly by the trends in the Curie-Weiss temperature with doping. Moreover, the ambiguous broadened transition edge at lower temperatures in off-stoichiometric \YTOd samples could interfere with low-temperature measurements of the ground state. We hope that having access to more pure single crystals of \YTO will advance the ability to accurately probe and understand this intriguing material. 

\section*{Acknowledgments}
The Institute of Quantum Matter is supported by Department of Energy (DOE), Office of Basic Energy Sciences, Division of Materials Sciences and Engineering under award DE-FG02-08ER46544. The authors acknowledge helpful input from and conversations with Collin Broholm.

\clearpage
\onecolumngrid
\newpage
\appendix*
\section{}

\begin{verbatim}

\end{verbatim}
	\renewcommand\arraystretch{1.3}
	\begin{table*}[!ht]
		\caption{A comprehensive list of \YTO low-temperature heat capacity peaks in the current literature.\cite{Physica_1969,Dalmas_2006,Yaouanc_2011,Ross_2011_PRB,Chang_2012,Ross_2012,DOrtenzio_2013,Chang_2014} Values are estimated off published images. The full width at half-max (FWHM) is a measure of the broadness of a peak.}
		\label{YTO:table:HC}
		\begin{tabular}{@{\extracolsep{\fill}}lllll}
		\hline \hline
		\textbf{Publication} & \textbf{Sample} & \textbf{C(J K$^{-1}$ [mol f.u.]$^{-1}$)} & \textbf{T(mK)} & \textbf{FWHM(mK)}\\
		\hline
		Ross, 2011\cite{Ross_2011_PRB}&powder&370&268&7\\
		this work&single crystal&87&268&10\\
		D'Ortenzio, 2013\cite{DOrtenzio_2013}&polycrystal&73&265&14\\
		Chang, 2014\cite{Chang_2014}&powder&26&266&23\\
		Bl{\"o}te, 1969\cite{Physica_1969}&powder&20&210&28\\
		Ross, 2012\cite{Ross_2012}&crystal: high-T feature&17&266&123\\
		this work&x = -.01&10&233&30\\
		Ross, 2012\cite{Ross_2012}&crystal annealed: high-T feature&9.2&267&39\\
		Dalmas de R{\'e}otier, 2006\cite{Dalmas_2006}&powder&9&265&10\\
		this work&x = 0&7.5&265&10\\
		Chang, 2014\cite{Chang_2014}&single crystal&6.8&197&28\\
		Chang, 2012\cite{Chang_2012}&single crystal A&6.8&197&20\\
		this work&x = +.01&6&190&35\\
		D'Ortenzio, 2013\cite{DOrtenzio_2013}&single crystal&4.3&186&31\\
		Ross, 2012\cite{Ross_2012}&crystal: low-T feature&4.1&195&23\\
		Ross, 2011\cite{Ross_2011_PRB}&single crystal B: high-T feature&4.0&268&21\\
		Ross, 2012\cite{Ross_2012}&crystal annealed: low-T feature&3.7&195&25\\
		Ross, 2011\cite{Ross_2011_PRB}&single crystal A&3.1&178&38\\
		Ross, 2011\cite{Ross_2011_PRB}&single crystal B: low-T feature&3.1&194&31\\
		this work&x = +.02&2.7&160&40\\
		Chang, 2012\cite{Chang_2012}&single crystal B&2.4&172&40\\
		Yaouanc, 2011\cite{Yaouanc_2011}&crystal \# 2 as grown&1.9&164&55\\
		Yaouanc, 2011\cite{Yaouanc_2011}&crystal \# 1 as grown&1.8&177&52\\
		Yaouanc, 2011\cite{Yaouanc_2011}&crystal \# 2 with heat treatment&1.8&163&41\\
		Chang, 2012\cite{Chang_2012}&single crystal C&n/a&n/a&n/a\\
		this work&x = +0.08&n/a&n/a&n/a\\
		\hline \hline
		\end{tabular}
	\end{table*}
	
\begin{figure*}
\centering
\includegraphics[width=\textwidth]{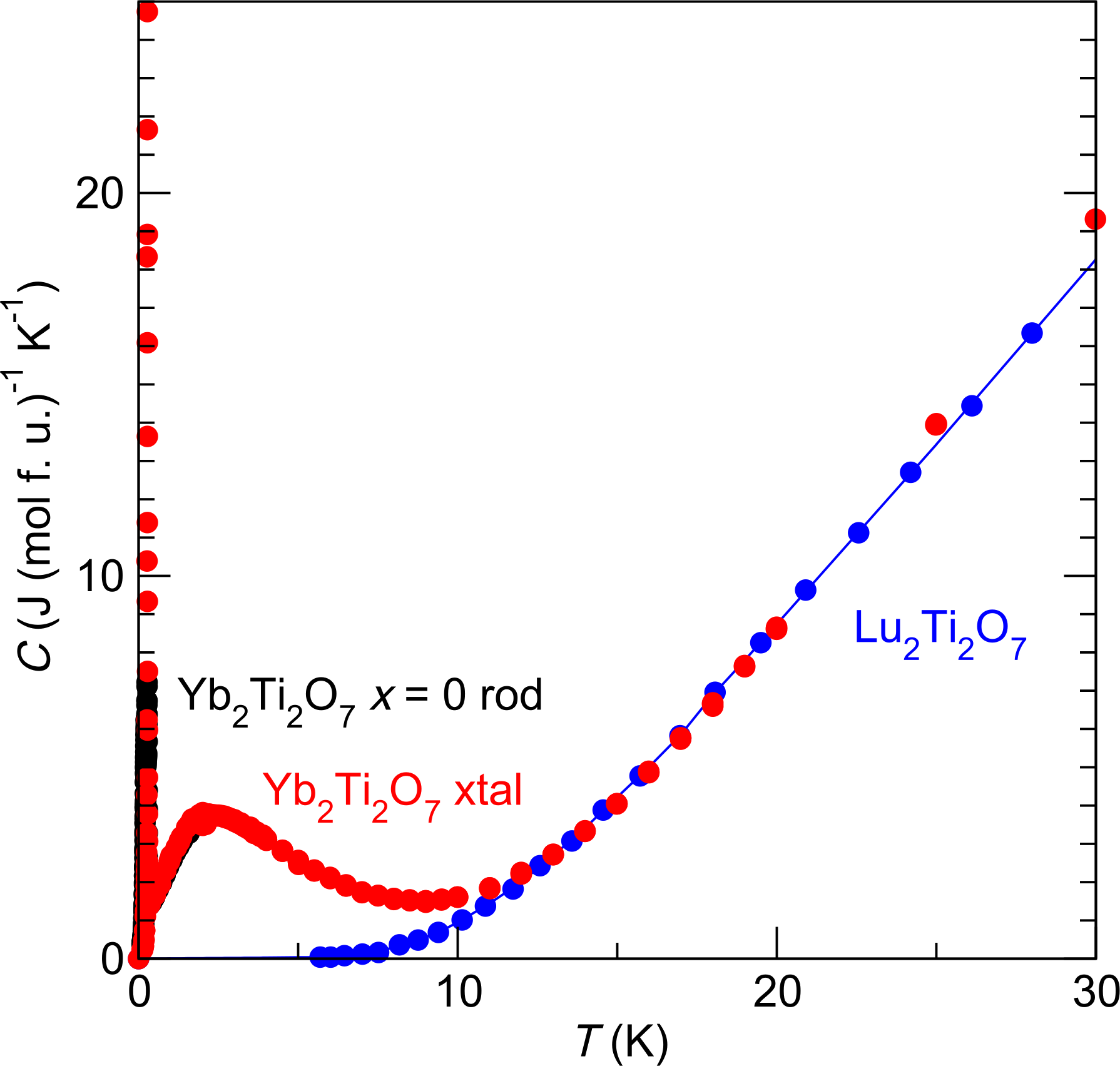}
\caption{The heat capacity of our \YTO single crystal (red) and $x = 0$ (black) sintered polycrystalline rod is shown with that of Lu$_2$Ti$_2$O$_7$\cite{FZGrowth_2013} (blue). Data are shown as circles; the blue line is the Lu$_2$Ti$_2$O$_7$ data extrapolated. Subtracting the extrapolated Lu$_2$Ti$_2$O$_7$ data from the \YTO data yields the magnetic heat capacity of \YTO used in the paper. (The low-temperature peak extends beyond the scale of this graph and is not shown in its entirety.)}
\label{fig:HCa}
\end{figure*}

\end{document}